\documentclass[structabstract]{aa}
\usepackage{graphicx}
\usepackage{epsfig}
\usepackage{color}
\usepackage{txfonts}
\usepackage{natbib}

\begin{document}

\title{Millimagnitude Photometry for Transiting Extrasolar
Planetary Candidates. V. Follow-up of 30 OGLE Transits.
New Candidates\thanks{Based on observations collected with the
Very Large Telescope at Paranal Observatory (ESO Programme
075.C-0427(A), DM and JMF visiting observers).}
}

\titlerunning{Millimagnitude Photometry of 30 OGLE Transits}

\subtitle{}

\author{P. Pietrukowicz\inst{1,2}
        \and
        D. Minniti\inst{1,3}
        \and
        R.~F. D\'iaz\inst{4}
        \and
        J.~M. Fern\'andez\inst{1,5}
        \and
        M. Zoccali\inst{1}
        \and
        W. Gieren\inst{6}
        \and \\
        G. Pietrzy\'nski\inst{6,7}
        \and
        M.~T. Ru\'iz \inst{8}
        \and
        A. Udalski\inst{7}
        \and
        T. Szeifert\inst{9}
        \and
        M. Hempel\inst{1}
        }

\authorrunning{Pietrukowicz et al.}

\offprints{P. Pietrukowicz,\\ \email{pietruk@astro.puc.cl}}

\institute{Departamento de Astronom\'ia y Astrof\'isica,
Pontificia Universidad Cat\'olica de Chile, Av. Vicu\~na MacKenna 4860,
Casilla 306, Santiago 22, Chile
       \and
Nicolaus Copernicus Astronomical Center, ul. Bartycka 18, 00-716 Warszawa,
Poland
       \and
Vatican Observatory, Vatican City State V-00120, Italy
       \and
Instituto de Astronom\'ia y F\'isica del Espacio (CONICET-UBA), Buenos Aires, Argentina
       \and
Harvard-Smithsonian Center for Astrophysics, 60 Garden Street,
Cambridge, MA 02138, USA
       \and
Departamento de Astronom\'ia, Universidad de Concepci\'on, Casilla 160-C,
Concepci\'on, Chile
       \and
Warsaw University Observatory, Al. Ujazdowskie 4, 00-478 Warszawa, Poland
       \and
Departamento de Astronom\'ia, Universidad de Chile, Casilla 36-D, Santiago, Chile
       \and
European Southern Observatory, Casilla 19001, Vitacura, Santiago 19, Chile
}

\date{Received April, 2009; accepted October, 2009}

\abstract
{}
{
We used VLT/VIMOS images in the $V$ band to obtain light curves of
extrasolar planetary transits OGLE-TR-111 and OGLE-TR-113, and candidate
planetary transits: OGLE-TR-82, OGLE-TR-86, OGLE-TR-91, OGLE-TR-106,
OGLE-TR-109, OGLE-TR-110, OGLE-TR-159, OGLE-TR-167, OGLE-TR-170,
OGLE-TR-171.
}
{
Using difference imaging photometry, we were able to achieve
millimagnitude errors in the individual data points.
We present the analysis of the data and the light curves, by measuring
transit amplitudes and ephemerides, and by calculating geometrical
parameters for some of the systems\thanks{Photometry of the transiting
objects is available at the CDS via anonymous ftp to cdsarc.u-strasbg.fr}.
}
{
We observed 9~OGLE objects at the predicted transit moments. Two other
transits were shifted in time by a few hours. For another seven objects
we expected to observe transits during the VIMOS run, but they were not
detected.
}
{
The stars OGLE-TR-111 and OGLE-TR-113 are probably the only OGLE objects
in the observed sample to host planets, with the other objects being very
likely eclipsing binaries or multiple systems. In this paper we also
report on four new transiting candidates which we have found in the data.
}

\titlerunning{Millimagnitude Photometry for OGLE Transiting Extrasolar Planetary Candidates}
\authorrunning{P.Pietrukowicz et al.}

\maketitle

\section{Introduction}

The field of extrasolar planets is developing rapidly, producing
exciting results at an accelerated pace. The discovery of the first
extrasolar ``hot Jupiter'' around the nearby solar-type star 51~Peg
using precise radial velocity measurements \citep{may95}
spurred a number of discoveries. Chief among these was the
discovery of transits around the nearby solar-type star HD~209458
\citep{char00,hen00}. The success of the radial velocity studies also
boosted extrasolar planetary searches using other techniques such as
microlensing and transits. Currently more than 60 transiting extrasolar
planets are known\footnote{See http://exoplanets.eu/}. Many candidates
were discovered by the OGLE team who carried out systematic searches,
monitoring millions of stars along fields located in the Milky Way disk
\citep{uda02a,uda02b,uda03,pont08}. Of more than 200 transiting
candidates, already seven OGLE transits have been confirmed as being
due to planets: OGLE-TR-10 \citep{kon05}, OGLE-TR-56 \citep{kon03},
OGLE-TR-111 \citep{pont04}, OGLE-TR-113 \citep{bou04,kon04},
OGLE-TR-132 \citep{bou04}, OGLE-TR-182 \citep{pont08},
OGLE-TR-211 \citep{uda08}. Most of the other targets are eclipsing low-mass
stars or brown dwarfs, or due to blends of normal stars, triplets, etc.

Why such interest in observing transiting candidates? The radial
velocities give orbital parameters such as period, semi-major axis,
and projected mass ($M~{\rm sin}~i$).
The transits give not only the orbital parameters like
period and inclination, but also the planet sizes: the eclipse
amplitude is simply $(r/R)^2$. Thus, from combined radial velocities
and transits we know the mass of the planet without the inclination
ambiguity, and the radius, which gives a density. The few planets so
studied appear to be indeed inflated gaseous planets, i.e. ``hot Jupiters''.
One difficulty is that the derived planet radii are only good to 10-15\%.
More accurate transit photometry is needed to improve those estimates,
as argued by \cite{mou04}. For example the discoveries by \cite{pont05a,pont05b}
of planet-sized stars around OGLE-TR-106 and OGLE-TR-122,
help to constrain the models of planetary systems.
These planets are under intense irradiation, which inflates
their sizes, depending on their own orbital parameters and intrinsic
characteristics \citep{bar03,burr02}. In some way the OGLE planets
constitute the extreme cases, because of their very short periods.

We conducted a programme to monitor photometrically the OGLE transit
candidates, and here we present precise photometry for these transits.
Some of the objects, namely OGLE-TR-109, OGLE-TR-111, OGLE-TR-113 have
been already analyzed by \cite{fer06}, \cite{min07}, \cite{diaz07},
respectively. In this paper we present an analysis of all OGLE transits
in the VIMOS images. We also have searched for new transits.

Section~2 gives details on the observations, selection and properties
of the sample. Section~3 describes reductions of the data. In sections~4-8
we present the results obtained from the observed light curves
of the OGLE transits. Section~9 describes new transiting candidates we
have found in the data. Finally, section~10 states our main conclusions.

\section{The sample}

Our program was allocated 4 nights with VIMOS at the Unit Telescope 3 (UT3) of the
European Southern Observatory Very Large Telescope (ESO VLT) at Paranal Observatory
during the nights of April 9 to 12, 2005. All four nights were clear
throughout, with sub-arcsecond seeing during most of the time.

Before the run we prepared maps of the positions of the transit candidates
in the OGLE fields and computed OGLE transit ephemerides. The selection of fields
was based mostly on maximizing the number of interesting transiting candidates
to be monitored given the VIMOS field of view. VIMOS is an imager and
multi-object spectrograph \citep{lef03}. Its field of view consists of
four $7\arcmin \times 8\arcmin$ fields covered by the four CCD chips arranged
in a square pattern with a separation gap of about $2\arcmin$. The CCD size is
$2048 \times 2440$ pixels with a pixel size of $0\farcs205$.

We selected four fields in Carina, that contain (but are not centered
on) the following transit candidates: OGLE-TR-86, OGLE-TR-113, OGLE-TR-167,
and OGLE-TR-170. We will refer to these fields as F86, F113, F167, and F170,
respectively. Table~1 gives basic information on the monitored fields.
In Fig.~1 we show the map of field F113 with OGLE transit candidates included.

\begin{table*}[htb]
\centering
\caption{Basic information on observed fields with VIMOS.
Coordinates are given for the centers of the fields.}
\smallskip
{\small
\begin{tabular}{ccccccc}
VIMOS field  & Short & RA(2000.0) & Dec(2000.0)                   & $l$       & $b$       & Duration of observations \\
with transit & name  & [h:m:s]    & [$\degr$:$\arcmin$:$\arcsec$] & [$\degr$] & [$\degr$] & \\
\hline
OGLE-TR-86~~ & F86~~ & 10:58:37.19 & -61:31:29.4 & 289.905 & -1.540 & only first 8h of night 1 \\
OGLE-TR-113  & F113  & 10:52:56.00 & -61:28:15.0 & 289.269 & -1.783 & all four nights \\
OGLE-TR-167  & F167  & 13:31:36.00 & -64:04:15.0 & 307.306 & -1.541 & 2h of night 1 and all night 2 \\
OGLE-TR-170  & F170  & 13:14:17.60 & -64:44:21.0 & 305.368 & -1.976 & only nights 3 and 4 \\
\hline
\label{table1}
\end{tabular}}
\end{table*}

\begin{figure}
\centering
\includegraphics[angle=0,width=0.5\textwidth]{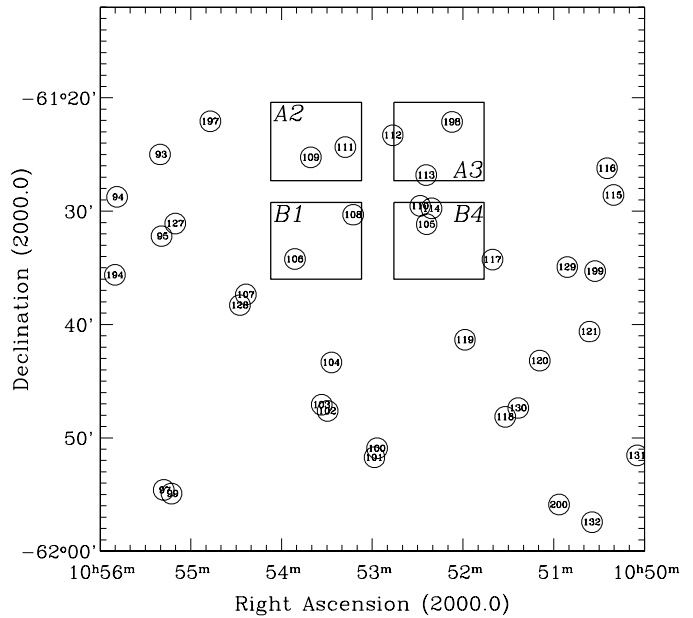}
\caption{Location of OGLE transit candidates in the area of OGLE-TR-113
or VIMOS field F113. VIMOS quadrants are marked and labeled as $A2$, $A3$,
$B1$ and $B4$.}
\label{mapTR113}
\end{figure}

We monitored these four fields, with a basic strategy to maximize the
observing efficiency trying to cover as many transit candidates as possible.
Two fields per night were observed alternatively, with three 15~sec
images acquired per field before moving to the other field. With the help
of the ESO Paranal staff, we reduced the nominal 9~min overhead between
two different field exposures to 90~sec. Typically we obtained 150-300 images
per night per field, resulting in well-sampled transits. All 32 OGLE transit
candidates located in the VIMOS fields are listed in Table~2. This table
indicates which candidates were expected to transit during our observations.

\begin{table*}[htb]
\centering
\caption{List of OGLE transiting candidates in the VIMOS fields.}
\smallskip
{\small
\begin{tabular}{ccccc|ccccc}
OGLE    & VIMOS & $P_{\rm OGLE}$ &  Events  & Were they & OGLE    & VIMOS & $P_{\rm OGLE}$ &  Events  & Were they \\
transit & field &       [d]      & expected & observed? & transit & field &       [d]      & expected & observed? \\
\hline
TR-81~~ & F86~~ & 3.2165(6)   & 0 &      -             & TR-111 & F113 & 4.0161(8)   & 1 & fully \\
TR-82~~ & F86~~ & 0.76416(14) & 0 & sinusoidal var     & TR-112 & F113 & 3.8790(8)   & 1 & star is saturated \\
TR-83~~ & F86~~ & 1.5992(3)   & 0 &      -             & TR-113 & F113 & 1.4325(3)   & 1 & fully \\
TR-84~~ & F86~~ & 3.1130(6)   & 1 &      no            & TR-114 & F113 & 1.7121(3)   & 1 & ?, worse seeing \\
TR-85~~ & F86~~ & 2.1146(4)   & 1 &      no            & TR-198 & F113 & 13.631(3)   & 0 &  -   \\
TR-86~~ & F86~~ & 2.7770(6)   & 1 & partially          & TR-159 & F167 & 2.1268(4)   & 1 & fully \\
TR-87~~ & F86~~ & 6.6067(13)  & 0 &      -             & TR-160 & F167 & 4.9018(10)  & 0 &  -   \\
TR-88~~ & F86~~ & 1.2501(3)   & 0 & star is saturated  & TR-161 & F167 & 2.7473(5)   & 0 &  -   \\
TR-91~~ & F86~~ & 1.5790(3)   & 1 & partially          & TR-162 & F167 & 3.7582(7)   & 1 & no   \\
TR-126  & F86~~ & 5.1108(10)  & 0 &      -             & TR-163 & F167 & 0.94621(18) & 0 &  -   \\
TR-192  & F86~~ & 5.4239(11)  & 0 &      -             & TR-164 & F167 & 2.6815(5)   & 0 &  -   \\
TR-105  & F113  & 3.0581(6)   & 0 &      -             & TR-166 & F167 & 5.2192(10)  & 0 &  -   \\
TR-106  & F113  & 2.5358(5)   & 1 &    fully           & TR-167 & F167 & 5.2610(10)  & 1 & partially \\
TR-108  & F113  & 4.1859(8)   & 1 &     no             & TR-170 & F170 & 4.1368(8)   & 1 & fully \\
TR-109  & F113  & 0.58909(12) & 4 & 3 events           & TR-171 & F170 & 2.0918(4)   & 1 & fully \\
TR-110  & F113  & 2.8486(6)   & 1 & partially          & TR-172 & F170 & 1.7932(4)   & 1 & no   \\
\hline
\label{table1}
\end{tabular}}
\end{table*}

We observed with the $V$ filter only, since the sampling rate did not allow us to use
two filters, and previous $I$-band observations are available from the OGLE database.
One of the main objectives of this work was to discard blends and binary stars
present among the transit candidates thanks to characteristic shape of transit events.
Also the light curves measured in the $V$-band can be compared with the $I$-band
OGLE light curves. While the $I$-band is more efficient for transit searches
\citep{pepp05}, the $V$-band shows the effects of limb darkening during the transit
better, and is more suitable for the modelling of the transit parameters.

\section{Photometry}

The bulk of the data acquired with VIMOS amounts to 82~GB. The periphery
of VIMOS images in each quadrant suffers from coma. Therefore, for our
analysis we cut a slightly smaller area of $1900 \times 2100$ pixels,
covering $7\farcm18 \times 6\farcm49$. The photometry was extracted with
the help of the {\it Difference Image Analysis Package} (DIAPL)\footnote{The
package is available at http://users.camk.edu.pl/pych/DIAPL/} written
by \cite{woz00} and recently modified by W. Pych. The package
is an implementation of the method developed by \cite{ala98}. To get better
quality of photometry we worked on $475 \times 525$ pixel subfields.

Reference frames were constructed by combining the 8-13 best individual images
(depending on the field and the quadrant). Profile photometry for
the reference frame was extracted with DAOPHOT/ALLSTAR \citep{stet87}.
These measurements were used to transform the light curves
from differential flux units into instrumental magnitudes, which later
were transformed to the standard $V$-band magnitudes by adding an offset
derived from $V$-band magnitudes of the transits in the fields \citep{diaz07,min07}.

The 15~sec exposure times saturate stars at $V\approx15.4$~mag
in the images under the best seeing. This affected two transit candidates:
OGLE-TR-88 and OGLE-TR-112.

The brightest planetary transit candidate monitored here is OGLE-TR-109,
with $I=14.99$ and $V=15.82$~mag, for which photometry with $\sigma_V=0.002$~mag
was obtained. The faintest candidate is OGLE-TR-108,
with $I=17.28$ and $V=18.73$~mag at $\sigma_V=0.009$~mag.

Due to the coma the photometric quality at the extreme corners
of the VIMOS fields is degraded. This affects the data for
OGLE-TR-126, and to a minor degree for OGLE-TR-87 and OGLE-TR-108.

\section{Results}

Twenty one transits were expected for 18 OGLE stars during the observations.
Figure~2 illustrates the transit times calculated for candidates in the field F113.
For object OGLE-TR-109 with the short period of $P=0.589127$~d,
four transits were expected. The object OGLE-TR-112 is saturated and no photometry
was obtained in this case. In total, we observed thirteen events in 11 stars.
Six transits were not detected and one is under question, due to worse seeing
at the end of night 4.

\begin{figure}
\centering
\includegraphics[angle=0,width=0.5\textwidth]{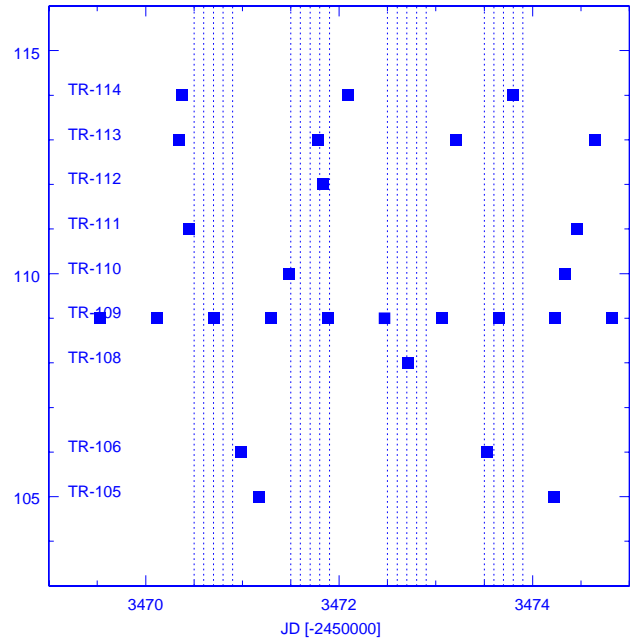}
\caption{Computed transit times for nine OGLE candidates in the field F113.}
\label{ephemeris}
\end{figure}

For 14 stars no transits were expected. However, even flat light curves
are useful to examine possible light curve modulations due to ellipsoidal
variability \citep{dra03,sir03}.

Table~2 also indicates whether we observed a full transit, a partial transit,
no transit, or sinusoidal variations. The relatively small number of expected
and detected transits in the fields F86 and F167 is the result
of the short period of observation of these two fields. The highest efficiency
is for F113, which is the only field observed for almost all 4 nights.

\section{The Significance of the Transits}

Because we will be dealing with potential unconfirmed transits or
false positives, it is necessary to find a quantitative way to
evaluate the significance of the individual transits observed.

The observed transits are well sampled in the $V$ band,
and the scatter is smaller than that of the OGLE transits.
There are typically $N_{\rm tr}=30-60$ points per transit in our data.
VIMOS transits are well sampled, allowing us to measure accurate amplitudes
as a difference between averages of the points outside and at the bottom of
the transits. In the case of OGLE, the significance of the transits is in part
judged by the number of transits detected (from a few to about 30).
In the case of the present study, we compute the signal-to-noise of the single,
well sampled transit. For a given photometric precision of a single measurement
of $\sigma_{\rm ph}$ and a transit depth $A$, this signal-to-noise transit is
$S/N=N_{\rm tr}^{1/2} A /\sigma_{\rm ph}$ \citep{gau05}. For the monitored transits we
find the range of $S/N$ to be 20-50 for typical $\sigma_{\rm ph}=0.004-0.005$~mag.
The error bars in the OGLE data are typically by 40-60\% larger than from
VIMOS.

We also computed the significance of the transits considering the
presence of correlated noise in the light curves, following the method
described in \cite{pont06}. Since we found that our light curves
exhibit moderate red noise, the values of the $S/N$ reported below for
different transits have been computed using this method.

It is also important to compute accurate mean times of transit
as well as to evaluate the transit timing errors,
for studies of multiplicity in these systems, as it has been done
in the case of OGLE-TR-111 by \cite{diaz08}.
The precision with which the mean transit time can be determined can
be estimated as: $\delta_{\rm tr}=\sigma_{\rm ph} t_{\rm tr}/(2 A \sqrt{N_{\rm tr}})$, where
$t_{\rm tr}$ is the transit duration, $N_{\rm tr}$ is the number of measurements
within transit \citep{deeg04}.

\section{Full OGLE Transits}

These objects allowed us to measure fundamental parameters of the systems
by applying analytical and empirical models that mimic transiting light curves.
In our work we used the method presented in \cite{man02}. For error
estimation we used the ``rosary-bead'' method fitting the light
curves with the downhill-simplex algorithm \citep{bou05,sou08,winn08}.
Below we discuss each of the observed objects.

\subsection{OGLE-TR-106}

In Fig.~3 we compare light curves for this object in the $I$ (OGLE)
and $V$ (VIMOS) bands. The transit occurred at the beginning of night~4 as
it was predicted. The shape of the event resembles a planetary transit,
but it is not! \cite{pont05a}, based on 8 spectra obtained with FLAMES
instrument at ESO VLT/UT2 telescope, show that this is an eclipsing binary,
where the secondary is an M dwarf of the mass of only $0.116 \pm 0.021$~$M_\odot$.
In Fig.~3 we also present our best fit to the VIMOS curve,
using linear limb-darkening coefficient for the $V$ band. This fit yields
ratios $r/R=0.145^{+0.007}_{-0.006}$, $a/R=13.3^{+1.1}_{-2.1}$,
the impact parameter $b=a/R~{\rm cos}~i=0.35^{+0.29}_{-0.35}$ at inclination
$i\approx90\degr$. The limb-darkening law parameter was fixed to $u=0.6$.
The estimated radius ratio is in agreement at $1\sigma$ level with the value
of $r/R=0.138\pm0.014$ derived from $R=1.31\pm0.09$ and $r=0.181\pm0.013$
\citep{pont05a}.

\begin{figure}
\centering
\includegraphics[angle=0,width=0.5\textwidth]{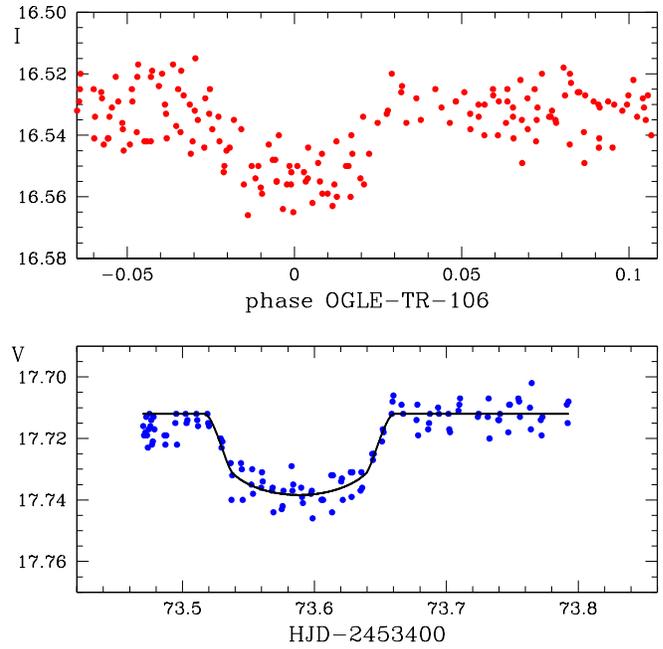}
\caption{Comparison of light curves for OGLE-TR-106 from OGLE (top panel)
and VIMOS (bottom panel). The magnitude scale and the time scale are
the same for both panels. Fit to the VIMOS data is presented.}
\label{lcTR106}
\end{figure}

\subsection{OGLE-TR-109}
This is an extreme case among the transiting candidates found by the
OGLE group because of early spectral type of the star (F0V), low
transit amplitude ($A_I\approx0.008$~mag), and very short period ($P=0.589127$~d).
Analysis of these photometric data by \cite{fer06}, and analysis of
high-resolution spectroscopic data by \cite{pont05a}, have left the nature of
the object undetermined. Two scenarios are possible: OGLE-TR-109 is either
a blend with a background eclipsing binary or a transiting planet.
However, the latter possibility seems to be less likely due to
the very short orbital period. Fig.~4 shows both OGLE and VIMOS light
curves. In comparison to the results published in \cite{fer06} the transit
$S/N$ is slightly better in this work (20 vs. 17), but still insufficient
to resolve more details in the light curve.

\begin{figure}
\centering
\includegraphics[angle=0,width=0.5\textwidth]{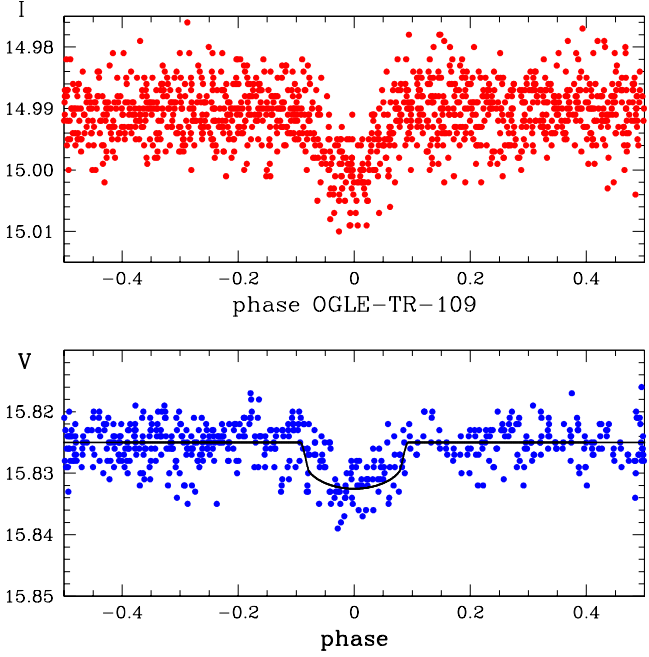}
\caption{Comparison of light curves for OGLE-TR-109 from OGLE (top)
and this work (bottom). The magnitude scale and the time scale are
the same for both panels. The best fit to the VIMOS data is presented.}
\label{lcTR109}
\end{figure}

\subsection{OGLE-TR-111}
This transit is caused by a hot Jupiter \citep{pont04}. It was analyzed
in detail by \cite{min07}. The VIMOS data allowed them to refine the planetary
radius, obtaining $r=1.01 \pm 0.06$~$R_J$. Our estimation of the radius ratios,
$r/R$=$0.1284^{+0.0066}_{-0.0033}$ and $a/R$=$12.31^{+0.74}_{-1.56}$,
is in excellent agreement with the published values
of $0.1245^{+0.0050}_{-0.0030}$ and $12.11^{+1.00}_{-1.39}$,
respectively. Recently, \cite{diaz08} found possible period
variations in this system which could be explained by the presence
of a perturbing planet with the mass of the Earth in an exterior orbit.
Fig.~5 presents our best fit to the VIMOS data.

\begin{figure}
\centering
\includegraphics[angle=0,width=0.5\textwidth]{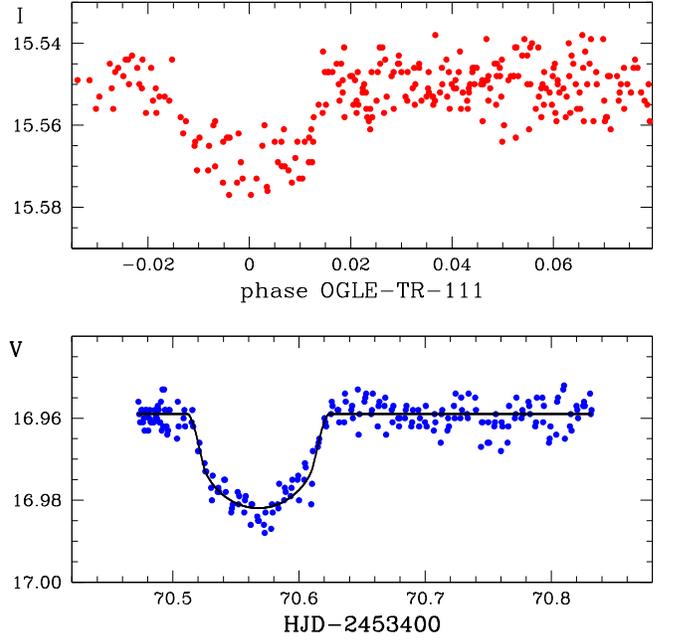}
\caption{Comparison of light curves for OGLE-TR-111 from OGLE (top)
and VIMOS (bottom). The magnitude scale and the time scale are
the same for both panels. The figure also shows the best transit
fit for this object.}
\label{lcTR111}
\end{figure}

\subsection{OGLE-TR-113}
This is another planetary transit \citep{bou04}.
It was expected to occur at the end of night~2 and it did.
\cite{diaz07} used the VIMOS data to obtain new estimates
for orbit parameters, radius and mean density of the planet OGLE-TR-113b.
The radius ratios we have found are almost identical to the values they
obtained: $r/R$=$0.1451^{+0.0064}_{-0.0022}$ vs. $0.1455\pm0.0083$,
and $a/R$=$6.49^{+0.10}_{-0.66}$ vs. $6.48\pm0.09$. The light curves
and the fit are shown in Fig.~6. Our new period estimation,
$P$=1.4324772(12)~d, is in excellent agreement with the value
of 1.4324757(13)~d derived by \cite{gill06} on the base of NTT/SUSI2
data taken almost at the same time (on 2005 Apr 3, 13).

\begin{figure}
\centering
\includegraphics[angle=0,width=0.5\textwidth]{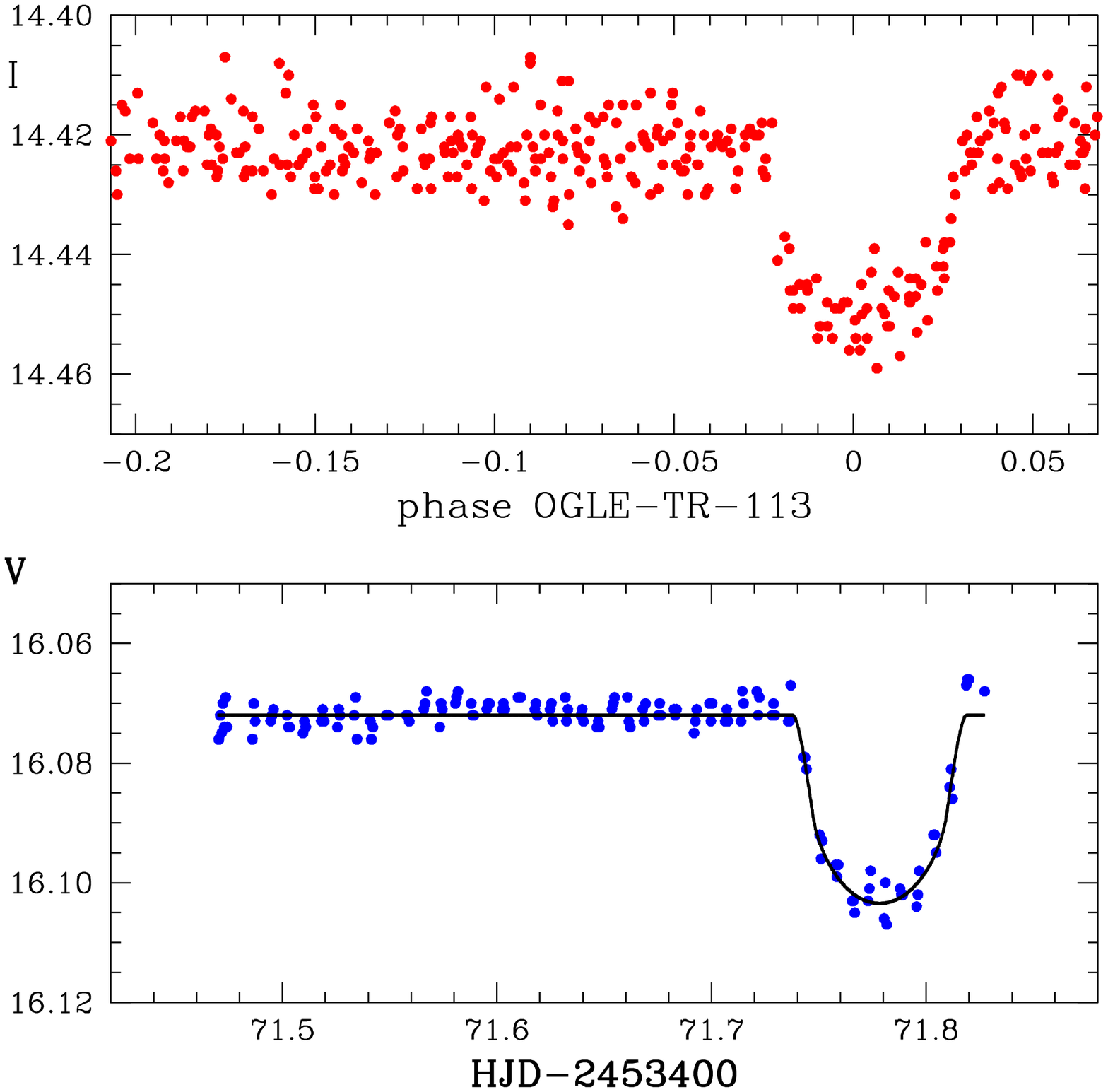}
\caption{Comparison of light curves for OGLE-TR-113 from OGLE in the
$I$ band (top panel) and VIMOS in the $V$ band (bottom panel).
The magnitude scale and the time scale are the same for both panels.
The solid line presents the best fit for this transit.}
\label{lcTR113}
\end{figure}

\subsection{OGLE-TR-159}
This object is located in the field of OGLE-TR-167. A transit
occurred in the middle of night~2, as it was expected from OGLE ephemeris.
The transit duration $t_{\rm tr}$ was approximately 3.2~hours, with about 45
data points ($\delta_{\rm tr}=1.3$~min). We measured an amplitude
$A_V=0.045 \pm 0.006$~mag with a transit signal-to-noise of $S/N\approx50$,
in agreement with $I$-band measurements at $1\sigma$ level:
$A_I=0.040 \pm 0.015$~mag. The transit portion of the light curve does
not show a flat bottom (see Fig.~7), indicating a large impact parameter for
this system. Following \cite{man02}, we obtained $r/R=0.41^{+0.03}_{-0.20}$
and inclination $i=74\degr$. We conclude that OGLE-TR-159 is a very likely
eclipsing binary.

\begin{figure}
\centering
\includegraphics[angle=0,width=0.5\textwidth]{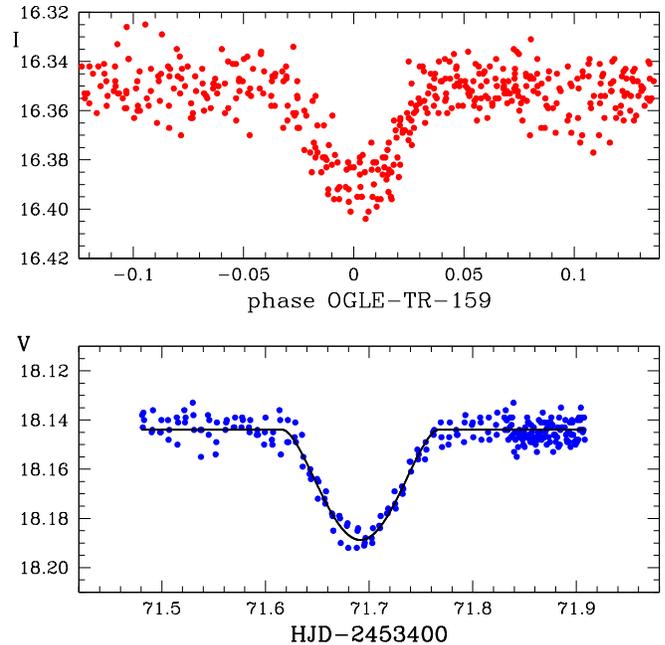}
\caption{Comparison of light curves for OGLE-TR-159 from OGLE (top)
and this work (bottom). The magnitude scale and the time scale are
the same for both panels. Fit to the VIMOS data is shown.}
\label{lcTR159}
\end{figure}

\subsection{OGLE-TR-170}
The transit OGLE-TR-170 occurred in the middle of night~3 and lasted for
$t_{\rm tr}\approx4.3$~h (see Fig.~8). There are more than 70 photometric points
($\delta_{\rm tr}=1.8$~min) within the transit. We measured an amplitude
$A_V=0.036 \pm 0.006$~mag with a transit signal-to-noise of
$S/N \approx 50$, again in agreement at $1\sigma$ level
with the $I$-band measurements: $A_I=0.030 \pm 0.015$~mag.
As for OGLE-TR-159, this transit does not show a flat bottom.
It is very likely that OGLE-TR-170 is an eclipsing system observed
during a grazing eclipse.

\begin{figure}
\centering
\includegraphics[angle=0,width=0.5\textwidth]{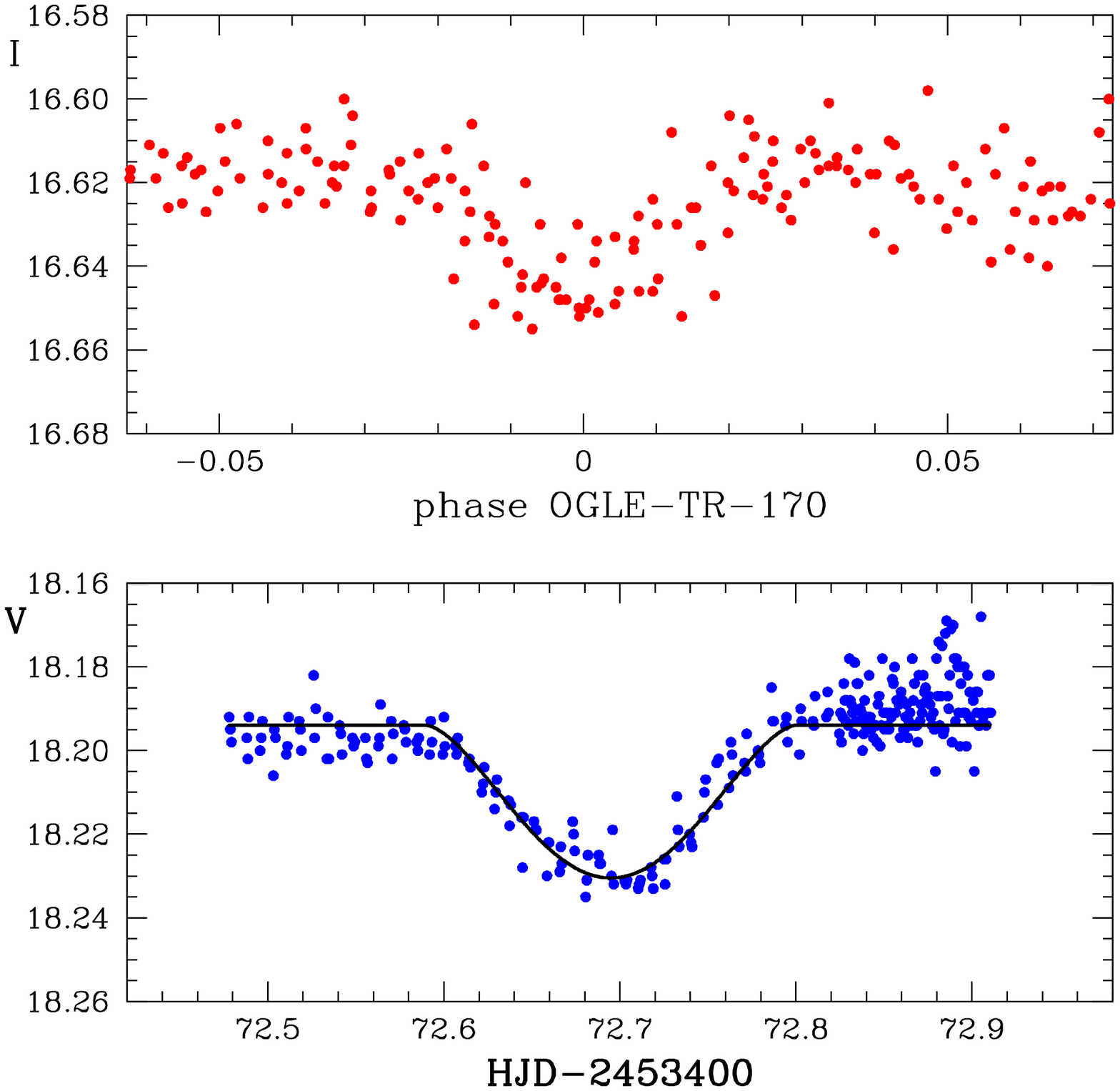}
\caption{Comparison of light curves for OGLE-TR-170 from OGLE (top)
and VIMOS (bottom). The magnitude scale and the time scale are
the same for both panels. The lower panel shows a fit of the event.}
\label{lcTR170}
\end{figure}

\subsection{OGLE-TR-171}
This transit occurred at the beginning of night~3 lasting about
2.5~hours. We measured an amplitude $A_V=0.032 \pm 0.009$~mag
with a transit signal-to-noise of $S/N\approx22$, what gave similar values
as in the $I$ band: $A_I=0.038 \pm 0.015$, $t_{\rm tr}=2.5$~h. The transit
seems to be asymmetric (Fig.~9). In both light curves, OGLE and VIMOS,
the minimum occurred earlier than the central moment of the event.
The asymmetry cannot be caused by a transiting planet, but could
indicate a matter flow between the components or the presence of a disk
around one of the stars.

\begin{figure}
\centering
\includegraphics[angle=0,width=0.5\textwidth]{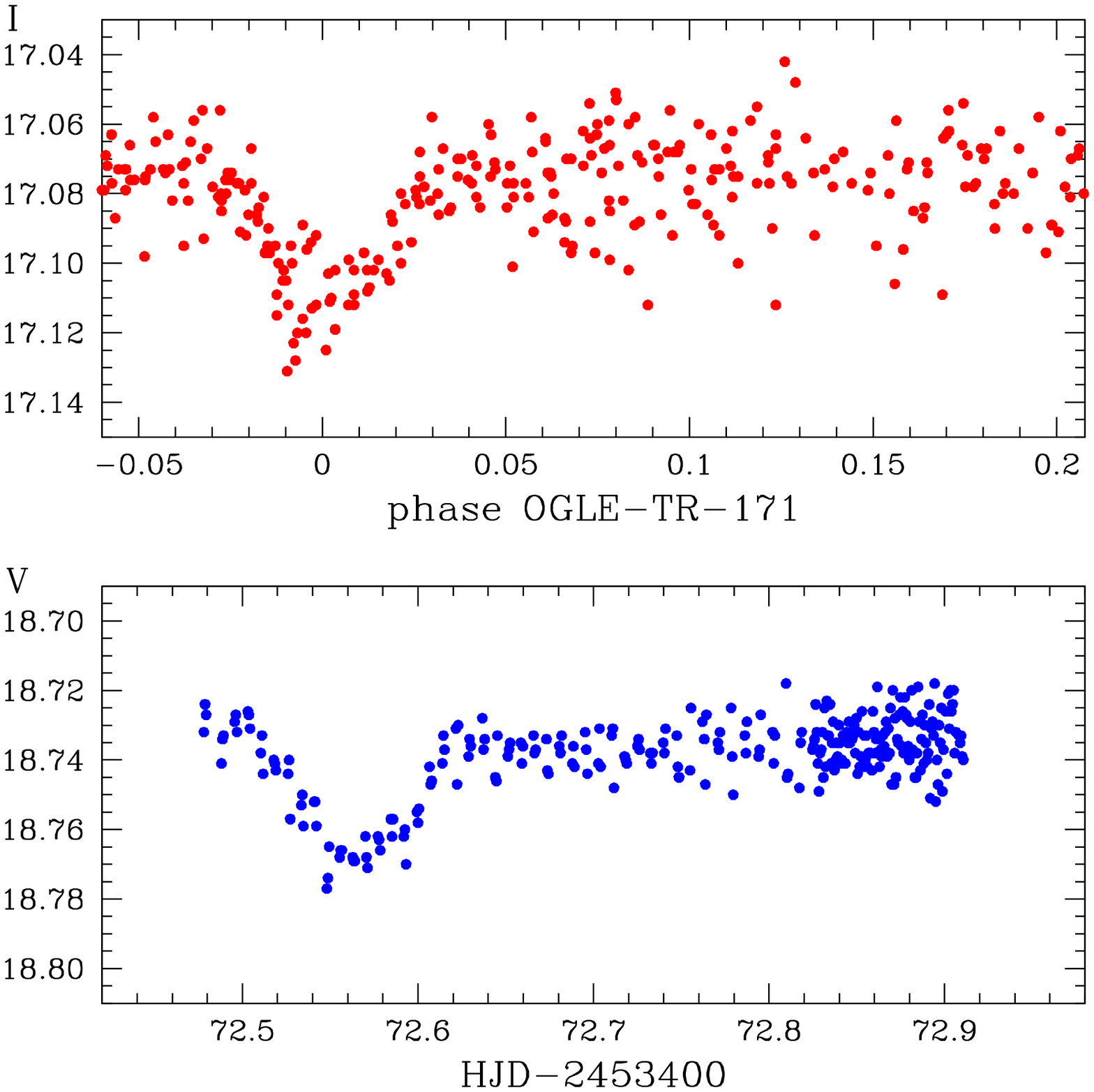}
\caption{Comparison of light curves for OGLE-TR-171 from OGLE (top)
and VIMOS data (bottom). The magnitude scale and the time scale are
the same for both panels. The shape of the transit is clearly asymmetric
and therefore no fit is given for this object.}
\label{lcTR171}
\end{figure}

\bigskip

In Table~3 we summarize geometrical parameters obtained for
six out of seven fully observed OGLE transits. In case of OGLE-TR-159
and OGLE-TR-170, for which V-shape eclipses were detected, the obtained
parameter distributions are bimodal, and therefore the given numbers
are not very representative of the distributions.
In the table we report median values at 68\% confidence level.

\begin{table}[htb]
\centering
\caption{Geometrical parameters determined for six systems. In the columns
we provide: ratio of the radii of the components $r/R$, ratio of the
radius of the orbit of the secondary to the radius of the primary,
impact parameter $b$, and linear limb-darkening coefficient $u$.
For objects OGLE-TR-109, OGLE-TR-111, and OGLE-TR-113 the limb-darkening
coefficient was taken from \cite{cla00}.}
{\small
\begin{tabular}{ccccc}
Transit      &         $r/R$                &          $a/R$          &         $b$            &  $u$ \\
\hline
             &                              &                         &                        & \\
\smallskip
\smallskip
OGLE-TR-106  & $0.1454^{+0.0071}_{-0.0060}$ &  $13.3^{+1.1}_{-2.1}$   & $0.35^{+0.29}_{-0.35}$ & 0.6 \\
\smallskip
\smallskip
OGLE-TR-109  & $0.0771^{+0.0037}_{-0.0029}$ &  $1.88^{+0.12}_{-0.25}$ & $0.17^{+0.48}_{-0.16}$ & 0.597$^*$ \\
\smallskip
\smallskip
OGLE-TR-111  & $0.1284^{+0.0066}_{-0.0033}$ & $12.31^{+0.74}_{-1.56}$ & $0.31^{+0.25}_{-0.31}$ & 0.768$^*$ \\
\smallskip
\smallskip
OGLE-TR-113  & $0.1451^{+0.0064}_{-0.0022}$ & $6.49^{+0.10}_{-0.66}$  & $0.11^{+0.35}_{-0.09}$ & 0.780$^*$ \\
\smallskip
\smallskip
OGLE-TR-159  &    $0.41^{+0.03}_{-0.20}$    & $4.18^{+0.24}_{-0.15}$  & $1.11^{+0.05}_{-0.15}$ & 0.6 \\
\smallskip
\smallskip
OGLE-TR-170  &    $0.23^{+0.19}_{-0.05}$    & $2.96^{+0.19}_{-0.10}$  & $0.91^{+0.26}_{-0.12}$ & 0.6 \\
\hline
\label{table4}
\end{tabular}}
\end{table}

\section{Partial OGLE Transits}

\subsection{OGLE-TR-86}
This transit occurred at the end of night~1 as it was expected from
OGLE data. We got only the ingress phase (see Fig.~10), so we could not
measure the flatteness of the light curve during the event.

\begin{figure}
\centering
\includegraphics[angle=0,width=0.5\textwidth]{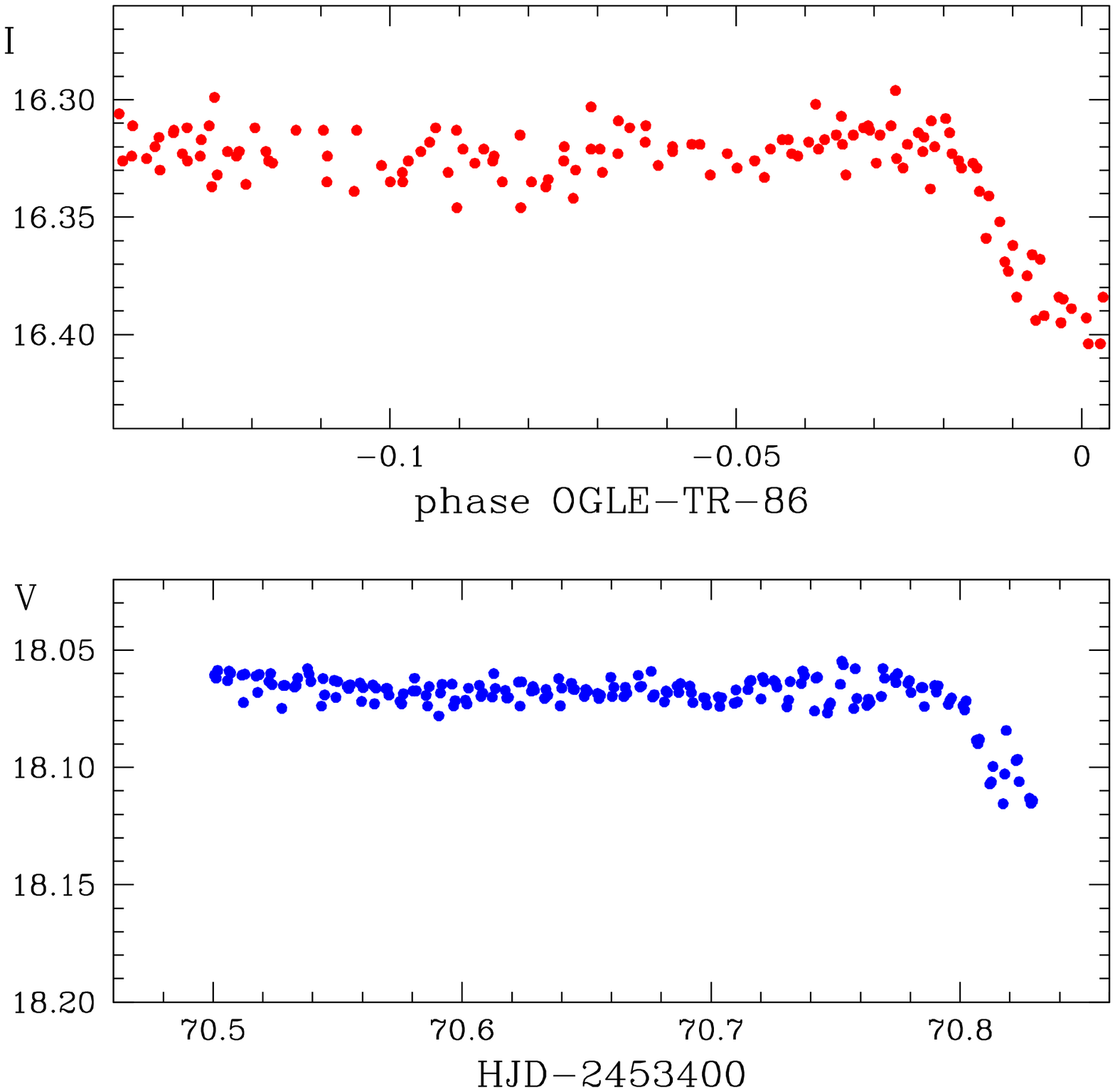}
\caption{Comparison of light curves for OGLE-TR-86 from OGLE (top)
and this work (bottom). The magnitude scale and the time scale are
the same for both panels.}
\label{lcTR86}
\end{figure}

\subsection{OGLE-TR-91}
The transit OGLE-TR-91 was expected in the middle of night~1, but it occurred
at the beginning of that night. The transit portion of the light curve
(Fig.~11) shows a flat bottom, indicating a relatively small impact parameter
for this system. We measured an amplitude $A_V=0.037 \pm 0.004$~mag,
in agreement with OGLE measurements: $A_I=0.043 \pm 0.015$~mag.
However, trapezium-like shape of the event with long ingress and egress
rather rules out planetary nature of the transit. A light curve modulation,
probably due to ellipsoidal variability, is also seen.

\begin{figure}
\centering
\includegraphics[angle=0,width=0.5\textwidth]{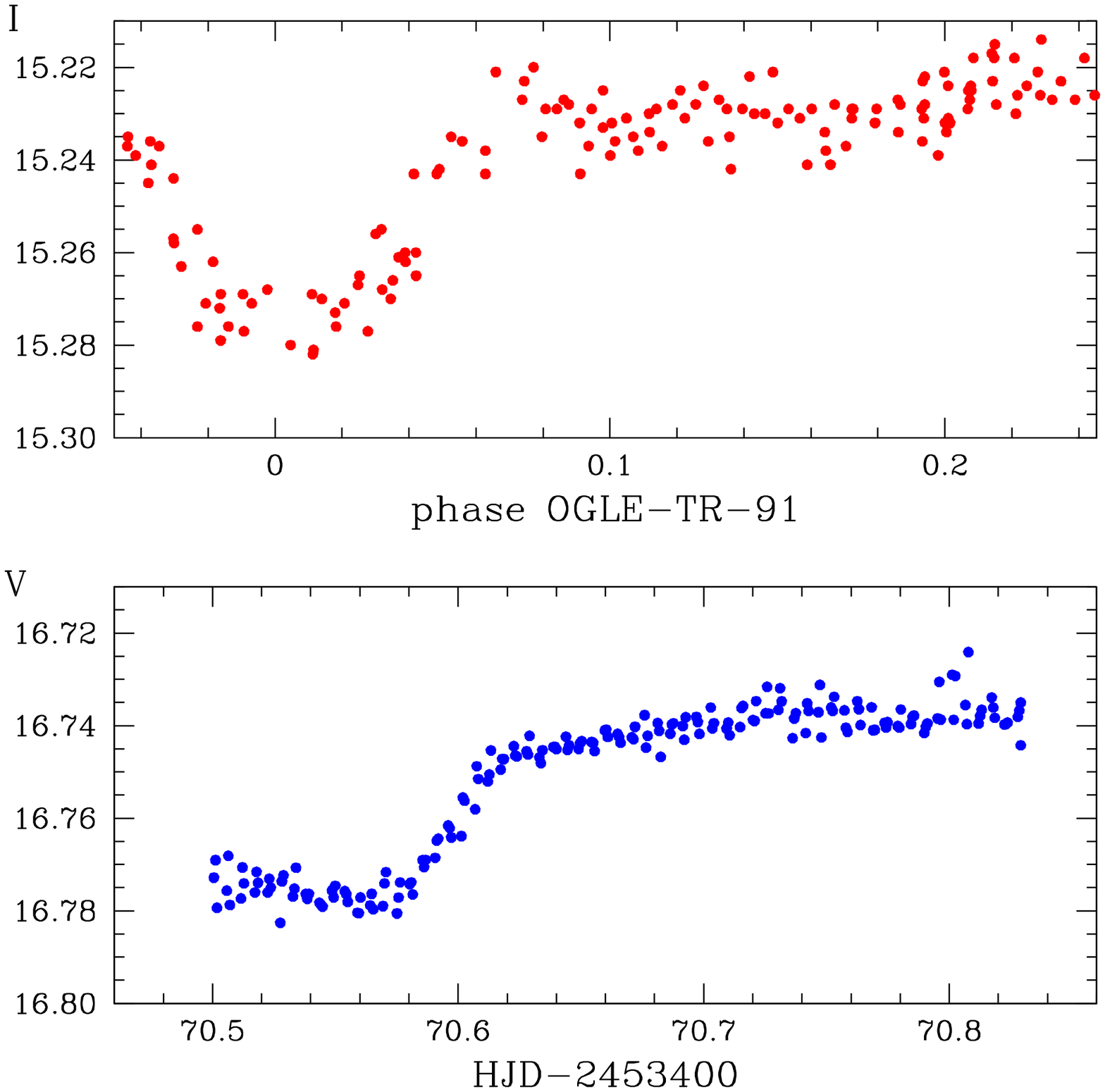}
\caption{Comparison of light curves for OGLE-TR-91 from OGLE in the $I$ band
(top) and VIMOS in the $V$ band (bottom). The magnitude scale and the time
scale are the same for both panels.}
\label{lcTR91}
\end{figure}

\subsection{OGLE-TR-110}
This object is located in the field F113. We observed only the egress
of the transit on night~2 (Fig.~12). Good agreement with the prediction
time indicates good estimation of the period from OGLE data.
Slow egress (0.03~mag in 1.5~h) clearly rules out planetary nature
of the event. This was confirmed by \cite{pont05a} who found large
radial velocity difference in two sets of spectral lines, like in
a binary.

\begin{figure}
\centering
\includegraphics[angle=0,width=0.5\textwidth]{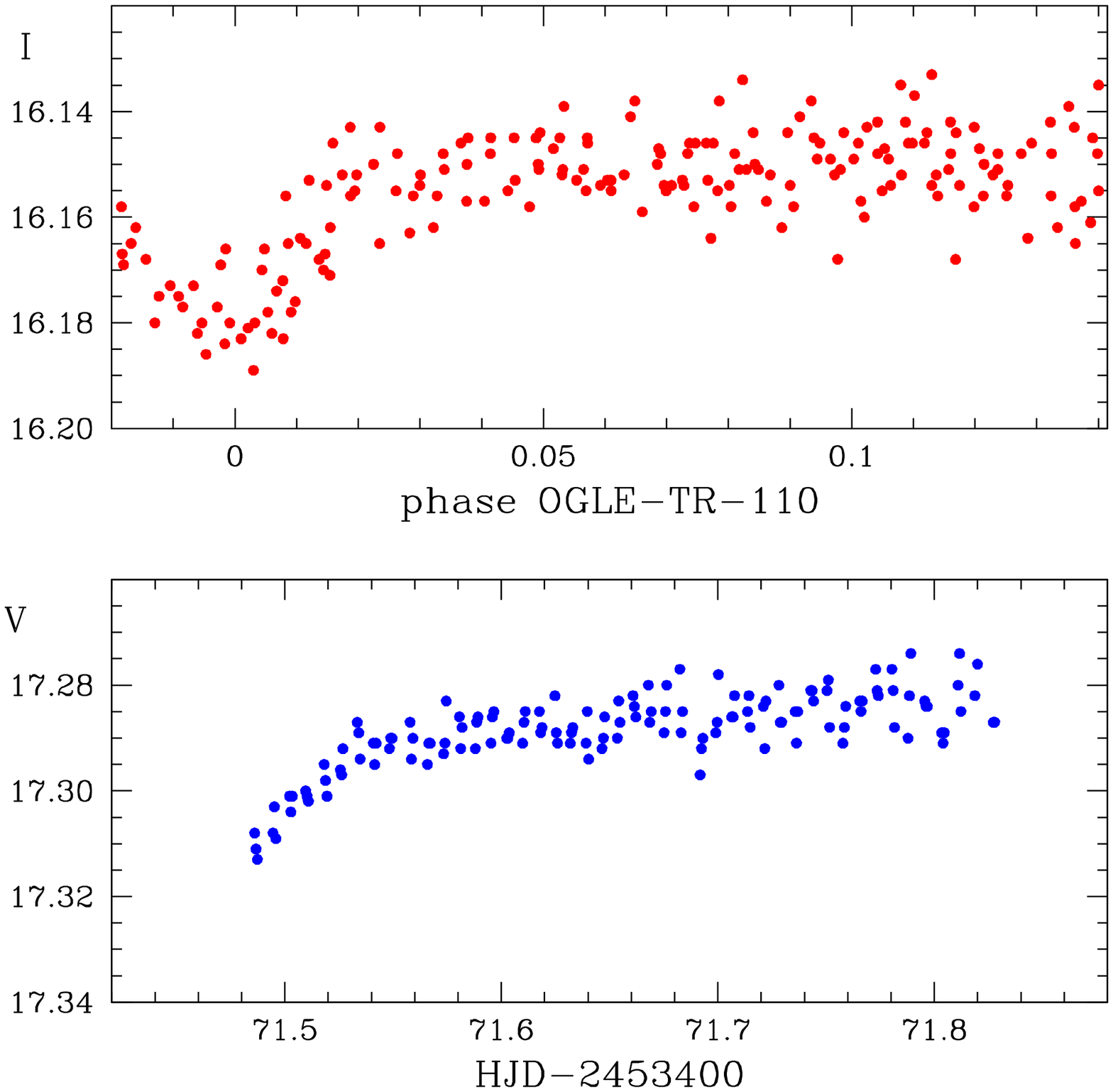}
\caption{Comparison of light curves for OGLE-TR-110 from OGLE (top)
and this work (bottom). The magnitude scale and the time scale are
the same for both panels.}
\label{lcTR110}
\end{figure}

\subsection{OGLE-TR-167}
This transit occurred at the end of night~2 as it was expected from
OGLE data. As for OGLE-TR-86, we only got the beginning of the
transit of OGLE-TR-167 (Fig.~13), so we couldn't measure the
flatteness of the light curve during the event.

\begin{figure}
\centering
\includegraphics[angle=0,width=0.5\textwidth]{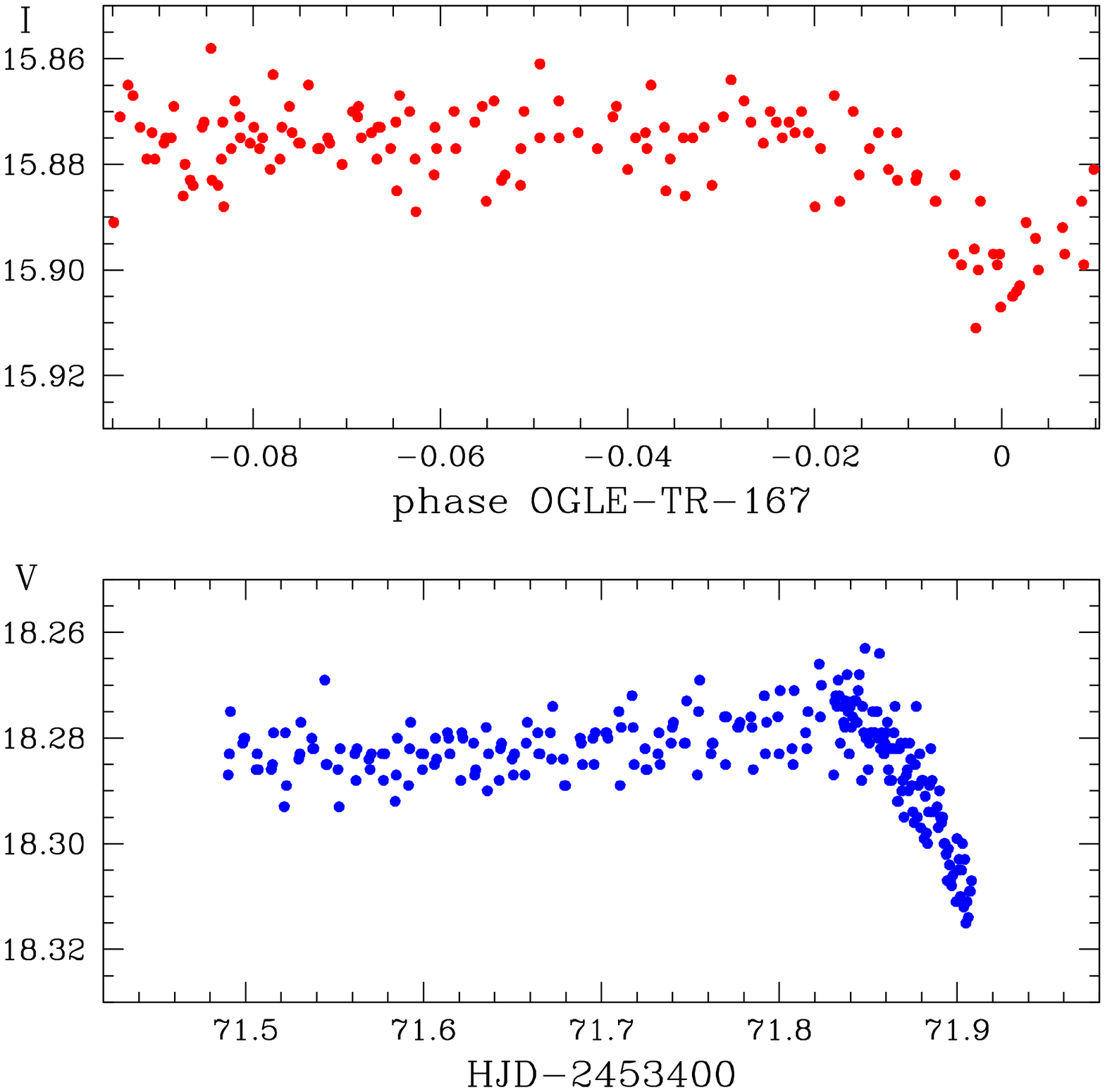}
\caption{Comparison of light curves for OGLE-TR-167 from OGLE (top)
and this work (bottom). The magnitude scale and the time scale are
the same for both panels.}
\label{lcTR167}
\end{figure}

\bigskip

For all partial transits but OGLE-TR-91 only $V$-band amplitude
lower limits could be measured. For the object OGLE-TR-91,
which is the only partial transit with a flat bottom, we estimated
the central moment of the event. For the other three objects their
periods were improved using either ingress or egress moments.
From the shape of the observed partial transits we conclude that none
of them was caused by a planet.

\begin{figure}
\centering
\includegraphics[angle=0,width=0.5\textwidth]{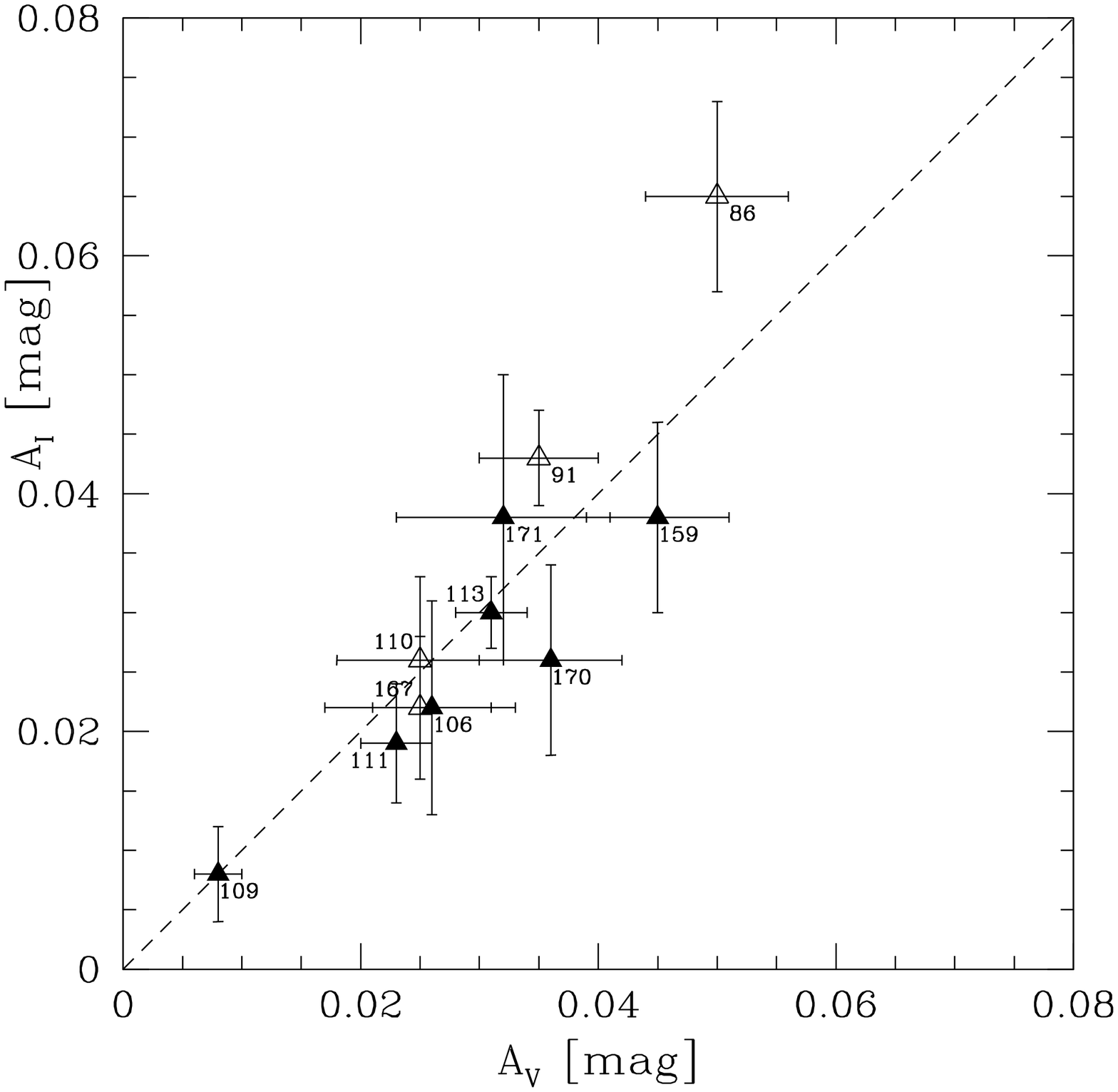}
\caption{$I$-band amplitudes {\it vs.} $V$-band amplitudes of the transits.
Fully and partially observed transits are marked with full and open triangles,
respectively.}
\label{amplitudes}
\end{figure}

In Table~4 we summarize photometric data on eleven objects with observed transits.
The measured $V$-band amplitudes of full transits are generally larger than the
amplitudes measured in the $I$ band (see Fig.~14). This agrees with the fact
that stellar limb darkening in the $V$ band should be shallower at the edges but
deeper by 10-20\% in the central parts \citep{cla03}.

\begin{table*}[htb]
\centering
\caption{Photometric data on OGLE objects for which either full or partial
transits were observed with VIMOS. The last five columns give, respectively:
central moment in Heliocentric Julian Days, transit duration $t_{\rm tr}$,
number of measurements within transit $N_{\rm tr}$, mean transit time
precision $\delta_{\rm tr}$ (see text), number of epochs $E$ that have passed
since OGLE observations, new period estimation. Note the dramatic increase
in precision of the periods in comparison to the values estimated by OGLE
(Table~2). Also note that period variations in the object OGLE-TR-111
(marked with *) have been recently reported by \cite{diaz08}.}
\smallskip
{\small
\begin{tabular}{ccccccccccccc}
OGLE    &  $I$  & $A_I$ & $\sigma_I$ &  $V$  & $A_V$ & $\sigma_V$ & HJD$_0-2453400$ & $t_{\rm tr}$ & $N_{\rm tr}$ & $\delta_{\rm tr}$ & $E$ & New period \\
transit & [mag] & [mag] & [mag] & [mag] & [mag] &  [mag]   &               &  [h:m]   &          &   [m]   &      &   [d]    \\
\hline
TR-106  & 16.53 & 0.022 & 0.009 & 17.71 & 0.026 &   0.005  &  73.5885(5)   &   3:00   &    50    &   1.8   &  453 & 2.535994(2)     \\
TR-109  & 14.99 & 0.008 & 0.004 & 15.82 & 0.008 &   0.002  &  72.540(1)    &   2:05   &    40    &   2.3   & 1950 & 0.5891262(8)    \\
TR-111  & 15.55 & 0.019 & 0.005 & 16.96 & 0.023 &   0.003  &  70.5676(5)   &   2:40   &    60    &   1.1   &  284 & 4.014510(4)$^*$ \\
TR-113  & 14.42 & 0.030 & 0.003 & 16.07 & 0.031 &   0.003  &  71.7782(5)   &   1:55   &    30    &   0.7   &  801 & 1.4324772(12)   \\
TR-159  & 16.35 & 0.038 & 0.008 & 18.14 & 0.045 &   0.006  &  71.6917(5)   &   3:10   &    45    &   1.3   &  364 & 2.126770(3)     \\
TR-170  & 16.62 & 0.026 & 0.008 & 18.19 & 0.036 &   0.006  &  72.6949(5)   &   4:15   &    70    &   1.8   &  187 & 4.136697(5)     \\
TR-171  & 17.07 & 0.038 & 0.012 & 18.73 & 0.032 &   0.009  &  72.562(1)    &   2:30   &    40    &   2.4   &  371 & 2.091819(4)     \\
\hline
TR-86~~ & 16.32 & 0.065 & 0.008 & 18.07 & $>$0.050 & 0.006 &   70.85(1)    &    -     &     -    &    -    &  413 & 2.77702(2)      \\
TR-91~~ & 15.23 & 0.043 & 0.004 & 16.74 & 0.035    & 0.005 &   70.54(1)    &    -     &     -    &    -    &  726 & 1.57883(2)      \\
TR-110  & 16.15 & 0.026 & 0.007 & 17.28 & $>$0.025 & 0.007 &   71.45(1)    &    -     &     -    &    -    &  402 & 2.84852(2)      \\
TR-167  & 15.88 & 0.022 & 0.006 & 18.28 & $>$0.025 & 0.008 &   71.95(1)    &    -     &     -    &    -    &  147 & 5.26066(7)      \\
\hline
\label{table4}
\end{tabular}}
\end{table*}

\section{Absent OGLE Transits}

\subsection{OGLE-TR-82}
This star lies in the field F86, which was monitored for only 8~hours.
A transit was predicted to occur at the end of night~1, but the observations
of the field were finished about 2 hours before the dawn and no event
could not be recorded. The flat portion of the light curve (see Fig.~15)
was measured with a photometric precision of 0.003~mag, allowing us to detect
a sinusoidal variation of a full amplitude of 0.0035~mag and a period
of $\sim0.23$~d, which is different to the transiting period of
$P=0.76416$~d. Recently \cite{hoy07} have shown that the system
is a main-sequence binary blended with a background red giant.
The variability may come from the blend.

\begin{figure}
\centering
\includegraphics[angle=0,width=0.5\textwidth]{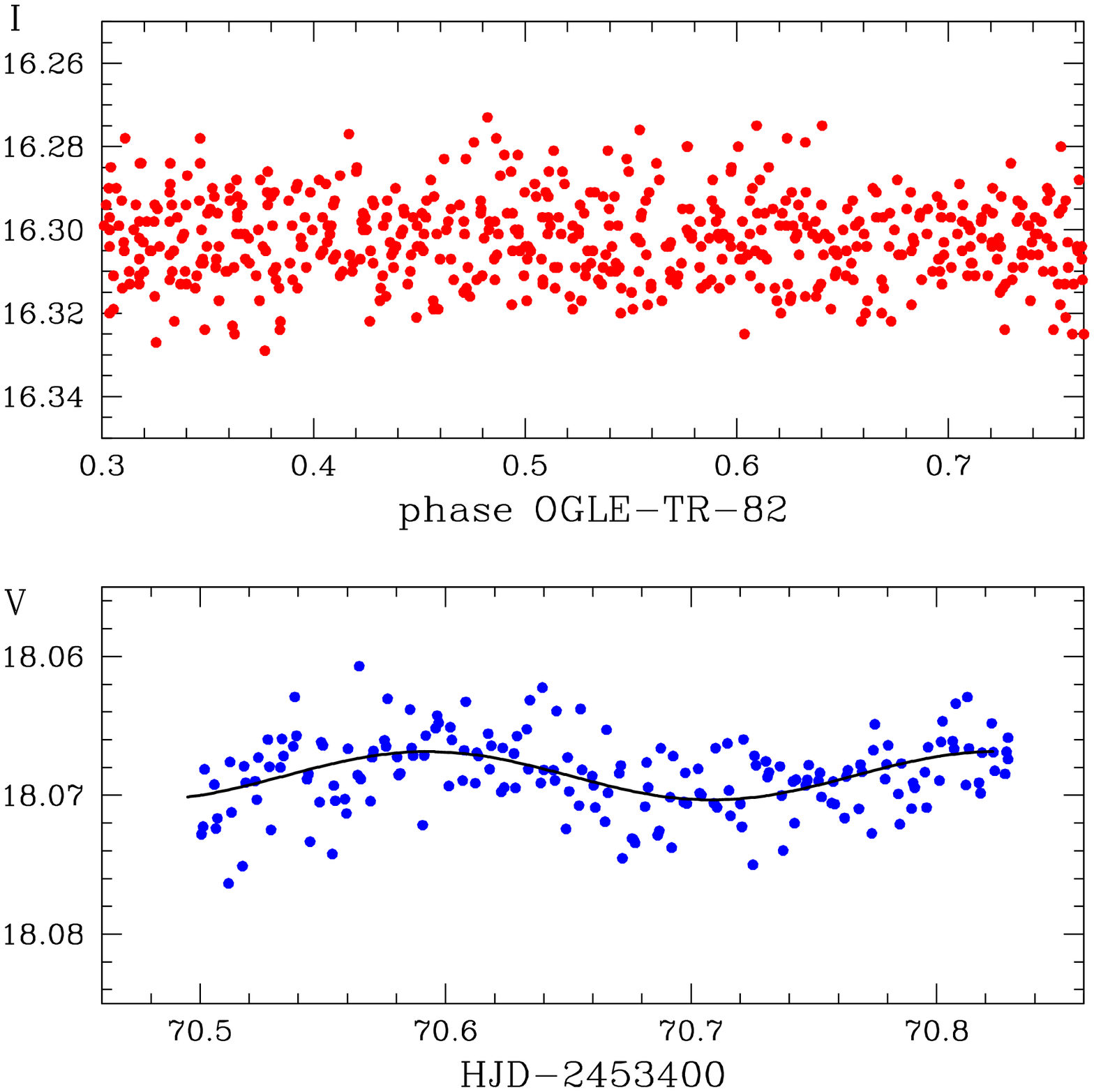}
\caption{Comparison of light curves for OGLE-TR-82 from OGLE (top)
and VIMOS (bottom). A sinusoidal fit to the VIMOS data is presented.}
\label{lcTR82}
\end{figure}

\subsection{OGLE-TR-84}
This object is located in the field F86. According to \cite{pont05a}
it is a likely eclipsing system. An eclipse was expected at the beginning
of night~1, but it was not detected.

\subsection{OGLE-TR-85}
Probably this is a triple system \citep{pont05a}. In this case we expected
to see a transit at the beginning of night~1, but nothing was observed.

\subsection{OGLE-TR-108}
This star is located in F113 field which was observed during all 4 nights.
Our ephemerides for OGLE-TR-108 predicted a transit in the middle of night~3,
and a secondary transit for night~1, but we did not detect anything.

\subsection{OGLE-TR-114}
A spectroscopic follow up made by \cite{pont05a} showed that this object
is a triple system. A transit was expected at the end of night~4,
but due to worse seeing there is no clear evidence for the event.

\subsection{OGLE-TR-162}
This object can be found in the field F167. In this case we hoped to observe
a transit in the middle of night~2, but nothing was detected.

\subsection{OGLE-TR-172}
The star is located in the VIMOS field F170. A transit expected to occur
in the middle of night~4 was not observed.

\bigskip

It is very likely that the seven transits mentioned above were not observed
due to two facts: short VIMOS observational run and lost of ephemerides for
the objects. This is clearly seen in the case of the transits from the
field F86 which was observed only for 8 hours. The time interval
between the OGLE and VIMOS observations was approximately three years.
Exact time uncertainties in the OGLE data are not given, but we assess
them to be below 0.0005~d. After the three years the errors of the
expected moments of the transits could have accumulated to a significant
fraction of the orbital period.

\section{New candidates}
We have also looked for new transit candidates in the VIMOS data. This has
been done in the framework of our variable search presented in \cite{pie09}.
Finding charts and light curves of four new candidates are shown
in Fig.~16 and Fig.~17, respectively. Table~5 gives the most important
observational facts on the transits. The first two objects are located
in the field F167 and the other two in the field F170. These are rather
faint stars with $17.8<V<19.7$~mag. The transits have amplitudes between
0.02 and 0.04~mag in $V$, and durations from about 2 to 5~hours.
All events were detected only once during our VIMOS run, therefore
there is no information on periods. We also note that for the transits~1-3
no events were detected in the OGLE-III data. The transit-4, due
to its depth and relatively long duration, is potentially the best candidate
for hosting a planet. This object lies outside OGLE-III fields and is
a good target for future surveys.

\begin{table*}[htb!]
\centering
\caption{Photometric information on four new transit candidates.}
\smallskip
{\small
\begin{tabular}{ccccccc}
New candidate & RA(2000.0) & Dec(2000.0) & $V$ & $A_V$ & HJD$_0-2453400$ & Duration \\
              &  [h:m:s]   & [$\degr$:$\arcmin$:$\arcsec$] & [mag] & [mag] & & [h:m] \\
\hline
transit-1 & 13:30:21.33 & -64:05:39.3 & 18.36 & 0.025 & 71.723 & 3:40 \\
transit-2 & 13:30:20.73 & -64:07:50.9 & 18.68 & 0.025 & 71.573 & 4:50 \\
transit-3 & 13:14:45.62 & -64:40:28.2 & 19.65 & 0.040 & 72.633 & 2:10 \\
transit-4 & 13:13:34.77 & -64:50:50.9 & 17.79 & 0.020 & 72.833 & 3:40 \\
\hline
\label{table5}
\end{tabular}}
\end{table*}

\begin{figure}[h!]
\centering
\includegraphics[angle=0,width=0.5\textwidth]{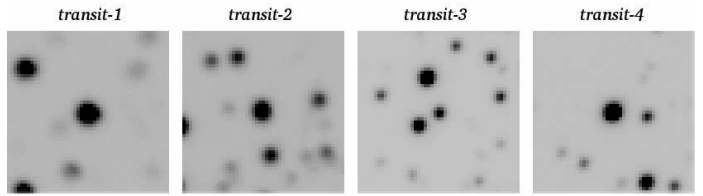}
\caption{Finding charts for the new transit candidates.
North is up and East to the left. The field of view is $10\arcsec$
on a side. The transiting stars lie exactly in the centers
of the charts.}
\label{charts}
\end{figure}

\begin{figure*}[htb]
\centering
\includegraphics[angle=0,width=1.0\textwidth]{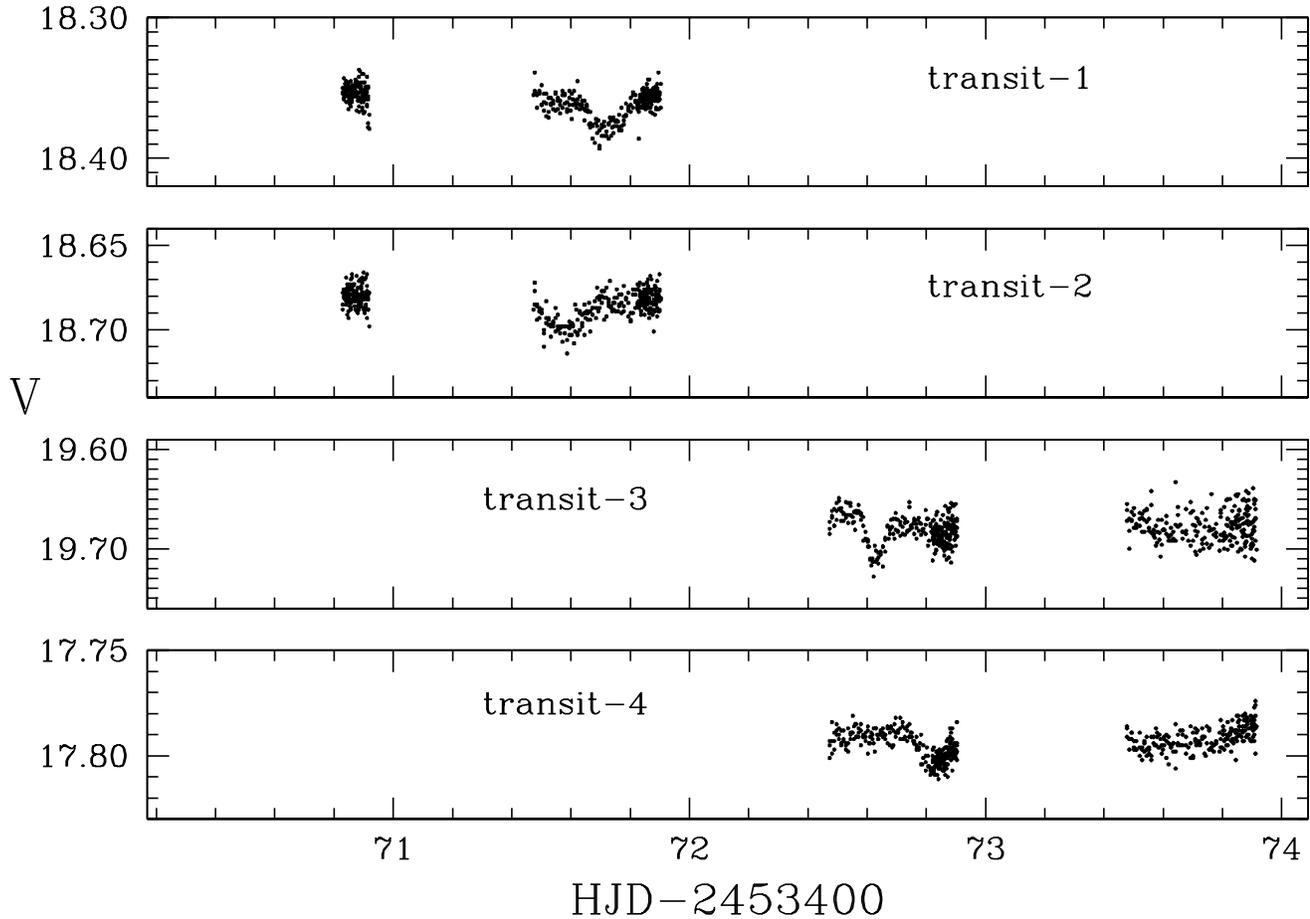}
\caption{New transit candidates found in VIMOS data.}
\label{new}
\end{figure*}

\section{Summary}

$V$-band images from VLT/VIMOS were used to obtain light curves of
extrasolar planetary transits: OGLE-TR-111, OGLE-TR-113, and
candidate planetary transits: OGLE-TR-109, OGLE-TR-159,
OGLE-TR-167, OGLE-TR-170, OGLE-TR-171. With difference imaging photometry
we were able to achieve millimagnitude errors in the individual data points.
The following seven OGLE transits were recorded as full events:
106, 109, 111, 113, 159, 170, 171. Four transits were detected
as partial events: 86, 91, 110, 167. All full and partial transits
but OGLE-TR-91 and OGLE-TR-109 were observed at the predicted transit
times. No transits were recorded for 19 objects.

Based on the shape of the obtained light curves and some results
from spectroscopic follow-up studies we show that the objects
OGLE-TR-111 and OGLE-TR-113 are probably the only OGLE stars
in the sample to host planets.

In the paper we also report on four new transiting candidates we have
found in the VIMOS data. One of the events, transit-4, with the duration
time of about 3.7~h and the amplitude of about 0.02~mag, is a particularly
good candidate for a planetary transit. Faintness of the object
(17.8~mag in $V$) may severely hamper spectroscopic verification.
However, all four candidates require photometric follow-up studies
to look for their periodic nature first.

\begin{acknowledgements}
PP, DM, JMF, GP, MZ, MTR, WG, MH are supported by FONDAP Center
for Astrophysics No. 15010003 and the BASAL Center for Astrophysics
and Associated Technologies. PP was also supported by the Foundation
for Polish Science through program MISTRZ and the Polish Ministry
of Science and Higher Education through the grant N N203 301335.
MZ acknowledges support by Proyecto FONDECYT Regular No. 1085278.
The OGLE project is partially supported by the Polish MNiSW grant
N20303032/4275. DM also thanks the John Simon Guggenheim Foundation.
AU acknowledges support from the grant ``Subsydium Profesorskie'' from the
Foundation for Polish Science. We thank the ESO staff at Paranal Observatory.

\end{acknowledgements}

\end{document}